%% file: main.tex
\documentclass[12pt]{article}	

\usepackage{amsmath}
\usepackage{amssymb}
\usepackage[preprint]{neurips_2022}
\usepackage{xspace}
\newcommand{\taskname}{\texttt{Mod-Cog}\xspace}
\newcommand{\yangtasks}{\texttt{20-Cog-tasks}\xspace}
\let\cite\citep

\usepackage[utf8]{inputenc} 
\usepackage[T1]{fontenc}    
\usepackage{hyperref}       
\usepackage{url}            
\usepackage{booktabs}       
\usepackage{amsfonts}       
\usepackage{nicefrac}       
\usepackage{microtype}      
\usepackage{xcolor}         
\usepackage{graphicx}

\title{Winning the lottery with neural connectivity constraints: faster learning across cognitive tasks with spatially constrained sparse RNNs}

\date{}

\author{%
   Mikail Khona$^*$ \\
   Physics \\
   MIT\\
   \texttt{mikail@mit.edu} \\
   \And
   Sarthak Chandra$^*$\\
   Brain and Cognitive Sciences\\
   MIT\\
   \texttt{sarthakc@mit.edu}\\
   \And
   Joy J. Ma\\
   Physics\\
   MIT\\
   \texttt{joym@mit.edu}\\
   \And
   Ila Rani Fiete \\
   Brain and Cognitive Sciences \\
   MIT \\
   \texttt{fiete@mit.edu} \\
}

\begin{document}

\maketitle

\begin{abstract}

Recurrent neural networks (RNNs) are often used to model circuits in the brain, and can solve a variety of difficult computational problems requiring memory, error-correction, or selection \cite{hopfield1982neural, maass2002real,maass2011liquid}. However, fully-connected RNNs contrast structurally with their biological counterparts, which are extremely sparse ($\sim 0.1$\%). 
Motivated by the neocortex, where neural connectivity is constrained by physical distance along cortical sheets and other synaptic wiring costs, we introduce locality masked RNNs (LM-RNNs) that utilize  task-agnostic predetermined graphs with sparsity as low as 4\%. We study LM-RNNs in a multitask learning setting relevant to cognitive systems neuroscience with a commonly used set of tasks, \yangtasks \cite{yang2019task}. 
We show through \textit{reductio ad absurdum} that \yangtasks can be solved by a small pool of separated autapses that we can mechanistically analyze and understand. Thus, these tasks fall short of the goal of inducing complex recurrent dynamics and modular structure in RNNs. We next contribute a new cognitive multi-task battery, \taskname, consisting of upto 132 tasks that expands by $\sim 7$-fold the number of tasks and task-complexity of \yangtasks. Importantly, while autapses can solve the simple \yangtasks, the expanded task-set requires richer neural architectures and continuous attractor dynamics. On these tasks, we show that LM-RNNs with an optimal sparsity result in faster training and better data-efficiency than fully connected networks.


\end{abstract}

\input{01_introduction}

\input{02_results}

\input{03_discussion}

\input{07_acknowledgements}

\clearpage

\input{04_references}

\clearpage

\clearpage

\input{06_appendix}

\end{document}

%% file: 01_introduction.tex
\section{Introduction}
Connectivity in biological neural networks are constrained by wiring-length costs, with a preference towards shorter synapses. The neocortex is effectively a two dimensional sheet which geometrically restricts physical distances between neurons ---
correspondingly, the spatial extent of connectivity in cortical circuits has been found to be significantly skewed towards shorter connections\cite{ercsey2013predictive,markov2013cortical,theodoni2022structural} (in particular connectivity extent appears to follow an exponential distribution). 

Inspired by such biological constraints, we modify the architecture of a vanilla fully-connected RNN in one particular fashion --- we construct a fixed sparse graph chosen by allowing local connections among neurons laid on a two-dimensional sheet. We then use this sparse graph for the RNN, by only training weights between nodes that correspond to edges on the graph, and setting all other weights to zero. We refer to an RNN in this set up as a `Locality Masked RNN' (LM-RNN).
Another motivation for our work comes from continuous attractor network models of grid cells \cite{khona2022smooth}, where it was shown analytically that fixed and local topographic connectivity encouraged the formation of discrete modules.

Exploiting such simple locality constraints in LM-RNNs, when applied to multitask regimes relevant to cognitive systems neuroscience, results in distinct advantages: these networks require far fewer parameters to train, there is no cost to performance relative to an unconstained network, and in fact learning is more rapid, sample-efficient and can acheive higher asymptotic performance. In contrast to other machine-learning approaches to network sparsification (see related work), we do not need any sophisticated pruning methods, any algorithms to construct and modify sparse skeletons nor any training data. 

We focus our results on multitasking regimes for RNNs, by training them to simultaneously learn many cognitive tasks. We expect our results on the improvements conferred by LM-RNNs to primarily apply in this multitask learning setting, where the recognition of modular structure across tasks is important for generalization and effective learning.

Our main contributions are summarized by:
\begin{itemize}
    \item We show that LM-RNNs can perform as well or better than dense networks, when accounting for the total number of nodes or the total number of synapses. Locality masking is thus an efficient prescription for choosing sparse subnetworks in a task agnostic and data independent fashion while still achieving high performance.
    \item We show that LM-RNNs reach this high performance faster and with lesser training than dense networks, indicating that sparse networks may be preferable to dense networks in memory and data-limited regimes for learning multiple tasks.
    \item We show that the tasks defined in  \cite{yang2019task} (an increasingly commonly used set of 20 tasks \cite{driscoll2022flexible,hummos2022thalamus,flesch2022modelling,marton2021efficient,riveland2022neural,elucidating20,duncker20organizing,masse2022rapid,kao2021natural} to study representations in networks performing many cognitive tasks) can be solved by a small pool of unconnected autapses, which is essentially a feedforward structure.
    \item Despite not have any lateral recurrent connections, the pool of autapses shows the existence of cell clusters according to the nodal task variance metric used in \cite{yang2019task}. Thus, our work reveals the limitations of using correlations or covariances between neurons to study recurrent mechanisms of computation. The shared inputs and input weights are a fundamental confound when using metrics such as nodal task variance used in \cite{yang2019task}. We thus highlight the need for creation of better metrics.
    \item We mechanistically study how this pool of unconnected autapses solves \yangtasks.
    \item We then introduce \taskname, a large battery consisting of upto 132 tasks inspired by cognitive science problems such as interval estimation, mental navigation and sequence generation which provides a useful setting to examine multitask learning and representation across tasks relevant to cognitive systems neuroscience.
    
\end{itemize}

\subsection{Related work}

Recent work has shown that effective pruning in recurrent networks can be done by biologically pruning plausible algorithms based on noise correlations between the presynaptic and postsynaptic neurons \cite{Moore20} that preserve the spectrum.
In vision neuroscience, recent works have studied cortical topography by using a pretrained vision frontend with a readout layer with an additional spatial loss to encourage spatially nearby cells to have correlated receptive fields \cite{finzi2022topographic,lee2020topographic,obeid2021wiring}. Other work has studied representations in RNNs trained to do simple tasks that have been embedded in 3-dimensional space \cite{}.

Several works in the realm of machine learning have tried to operationalize the idea of sparsity. These methods can be roughly categorized into 2 main classes:

\textbf{Dense-to-sparse}
Ref. \cite{han2015learning} experimentally showed that training followed by pruning and retraining can give sparse networks with no loss of accuracy and ignited an interest in pruning methods. Following this work, the lottery ticket hypothesis states that dense, randomly-initialized, feed-forward networks contain subnetworks (``winning tickets'') that --- when trained in isolation --- reach test accuracy comparable to the original network in a similar number of iterations  \cite{frankle2018lottery}. The best method to identify such winning tickets is Iterative Magnitude-based Pruning (IMP)  \cite{frankle2018lottery,frankle2019stabilizing}, which is computationally expensive and has to be run thoroughly for every different network. It has also been shown that parameters of the sparse initialization distribution and sign of weights at initialization are important factors \cite{zhou2019deconstructing} which determine winning tickets. Overall, iterative pruning and retraining methods involve 3 steps: (1) pre-training a dense all-to-all model, (2) pruning synapses based on some criteria, and (3) re-training the pruned model to improve performance. This cycle needs to be done atleast once and in many cases, multiple times, to get good performance. So this procedure requires at least the same training cost as training a dense model and often even more than that. In contrast, our proposed method for RNNs does not require seeing any data before pruning and trains over only the sparse remaining synapses; thus we do not require multiple cycles of pruning and training.

Other methods involving ways to encourage sparsity during the training process include $L_1$(Lasso) regularization \cite{wen2018learning}, $L_0$ regularization \cite{louizos2017learning, savarese2020winning} and pruning using dynamically varying thresholds \cite{narang2017exploring,kusupati2020soft}.
Unfortunately, all of the aforementioned methods require training the original dense network, in varying amounts, thus precluding the benefits that can be obtained by having a predetermined exact sparsity on the computation during training.  

A class of methods which do not involve training data like SynFlow \cite{tanaka20}, GraSP \cite{Wang2020Picking}, SNIP \cite{lee2018snip,Lee2020A} and FORCE \cite{de2020progressive} have been studied for only feedforward networks, while we study RNNs.

\textbf{Sparse-to-sparse}
Another line of work concerning sparse-to-sparse training is most relevant to our study. This involves using a sparse interaction graph which is used to mask gradient updates. Older works maintained a static graph  \cite{mocanu2016topological} and dealt only with feedforward networks but newer methods such as dynamic sparse training (DST) \cite{evci2020rigging, liu2021selfish} have been proposed for both feedforward networks and RNNs which dynamically improve the sparse graph and provide better performance. These methods generally involve changing the topology of the sparse graph during training. Ref. \cite{liu2021selfish} considers static Erdos-Renyi (ER) type sparse RNNs but for relatively denser values of sparsity (0.53 and 0.67) and more complex architectures like stacked LSTMs and Recurrent Highway Networks  \cite{zilly2017recurrent}. Here we explore more extreme values of sparsity ($\sim 5\%$ and below) and show that they are optimal in the context of multitask learning regimes.

Lastly, static sparse networks have also found common usage in reservoir computing architectures, where large sparse networks are preferred to fully-connected networks to increase heterogeneity across nodes and allow for ``richer'' dynamics \cite{jaeger2001echo,lukovsevivcius2009reservoir}.

%% file: 02_results.tex
\section{Results}





\subsection{Locality masked RNNs (LM-RNNs)}

\begin{figure}
    \centering
    \includegraphics[width = \textwidth]{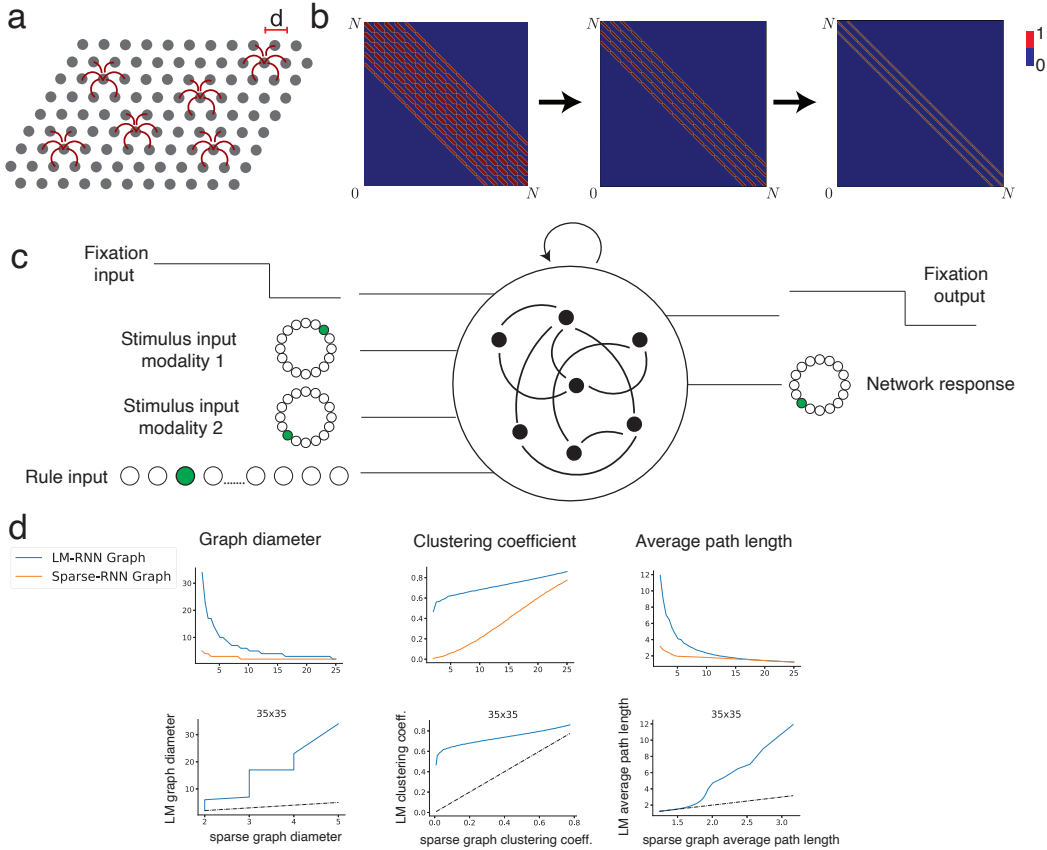}
    \caption{{\bf Local connectivity constraint (locality masking) and schematic of cognitive multitasking RNN.}(a) In LM-RNNs neurons are arranged on a two-dimensional sheet, with nonzero weights permitted for nodes up to a distance $\leq d$ apart. (b) Locality masking on a two-dimensional sheet can be treated as a sparse mask on the hidden-to-hidden weights of an RNN (c) Schematic of the RNN setup in the context of \yangtasks and \taskname: the network receives inputs encoding directions on two rings, a fixation signal and a rule input. The network is trained to output a fixation signal and a direction on a ring of output nodes. (d) A battery of graph theoretic metric to distinguish the connectivity graph of LM-RNNs from a random sparse (Erdos-Renyi) graph (left to right): Graph diameter, Clustering coefficient, Average path length.}
    \label{fig:Fig1}
\end{figure}

We restrict ourselves to simple RNNs for interpretability in the context of systems neuroscience. Our RNNs follow dynamics defined by:

\begin{align*}
    \mathbf{h}_{t+1} &= \phi( \mathbf{W}\mathbf{h}_{t} + \mathbf{W}_{in}\mathbf{u}_t + \mathbf{b}^h), \\
    \mathbf{o}_{t+1} &= \mathbf{W}_{out}\mathbf{h}_{t+1} +\mathbf{b}^{o}.
\end{align*}


Corresponding to the biological arrangement on neurons on a two-dimensional cortical sheet, we arrange the nodes of an RNN on the lattice points of a two-dimensional plane, as shown in Fig. \ref{fig:Fig1}a. Then, we constrain the weights for recurrent connections within the nodes of the RNN to be always zero for pairs nodes that lie at a Euclidean distance of larger than $d$. The training of the RNN then proceeds in the usual fashion by using back propogation of the loss to update the unconstrained weights. We refer to such an RNN as a \textbf{L}ocality \textbf{M}asked \textbf{RNN} (LM-RNN).

This constraint on the weights of the RNN is implemented through a graph $G$, whose nodes are the units of the RNN, and edges correspond to pairs of units with unconstrained weights, determined by the spatial distance $d$. Operationally, the adjacency matrix of the graph, $\mathbf{G}$, is point-wise multiplied with the interaction matrix $\mathbf{W}$ after each gradient step, effectively constraining which elements of the interaction matrix can be learnt by gradient descent:
\begin{equation}
    \mathbf{W}\leftarrow \mathbf{G}\otimes \mathbf{W}
\end{equation}
This graph adjacency matrix $\mathbf{G}$ is static and unchanged throughout training.

In effect, the two-dimensional LM-RNN consists of a sparse subgraph $G$ of the fully connected network, with each node connected to the $\sim \pi d^2$ nearest nodes to it. We posit that this sparse subgraph is like a ``winning lottery ticket'', such that when trained in isolation the LM-RNN achieves comparable performance to a fully recurrently trained network. Moreover, we will demonstrate that these winning lottery tickets perform better in a more data-efficient manner than fully connected counterparts with a similar number of nodes or a similar number of synapses. This approach can be implemented easily in all training frameworks and is agnostic to the specific optimization algorithm being used.

Constructing a sparse graph in this particular fashion is distinct from a sparse random graph. To distinguish between our locality masks and sparse random graphs (Erd\H{o}s-R\'enyi networks)  with the same number of edges, we borrow several metrics from graph theory. In particular we examine three metrics --- the average length of the shortest path between nodes, the graph diameter (i.e., the longest shortest path between two nodes) and correlation coefficient which (i.e., the density of mutually connected triplets of nodes)\cite{borner2007network}. Notably, these metrics are properties fundamental to the structure of the connectivity in a graph, and are invariant to re-lableing and permutation of nodes in the graph. As can be seen in Fig.\ref{fig:Fig1}d, there is a sharp contrast between the graph of an LM-RNN when compared with a random sparse RNN, with LM-RNNs having longer diameters and path lengths, but smaller clustering coefficients.

\subsection{LM-RNNs learn the `\yangtasks' more rapidly than fully-connected RNNs}
 
\begin{figure}
    \centering
    \includegraphics[width = \textwidth]{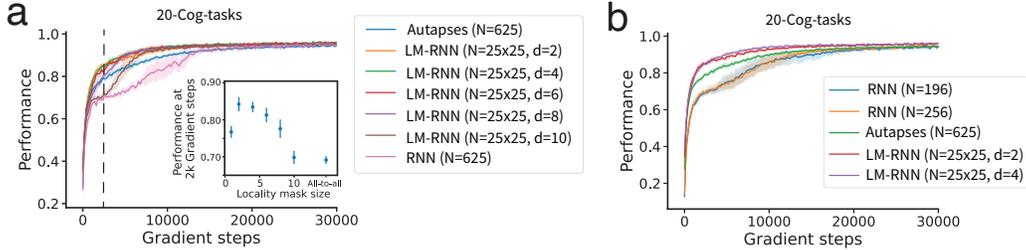}
    \caption{{\bf LM-RNNs learn \yangtasks faster than fully connected networks with matched neuron or synapse number:} (a) LM-RNNs for all choices of locality mask sizes, $d$, perform better than fully-connected RNNs with the same number of nodes. This includes $d=0$, i.e., a disconnected pool of autapses which also outperforms the fully-connected network. \textit{Inset}: performance at $2\times10^3$ gradient steps across different values of $d$ demonstrates a small optimal $d$ that leads to the best performance. (b) LM-RNNs also outperform fully-connected RNNs constructed with the same number of synapses. RNN($N=196$) and RNN($N=256$) have $5.2\times10^4$ and $8.3\times10^4$ parameters respectively; $N=625$ autapses, LM-RNN($N=25\times25$, $d=2$) and LM-RNN($N=25\times25$, $d=4$) have $4.4\times10^4$, $5.2\times10^4$ and $7.4\times10^4$ parameters respectively}
    \label{fig:Fig2}
\end{figure}

We apply LM-RNNs to a dataset used commonly in systems neuroscience, a set of 20 cognitive tasks introduced in \cite{yang2019task}, which we henceforth refer to as the \yangtasks. Each of these 20 tasks are constructed on the same input stimulus modalities --- two rings of input units are used that each support a single activity bump, encoding a one-dimensional circular variable (which could represent direction of motion, for example), which we represent with vectors $\mathbf{u}_t^\text{ring1}(\theta_1)$ and $\mathbf{u}_t^\text{ring2}(\theta_2)$. Along with the two rings, the input into the RNN also comprises two additional inputs: first, a one-hot encoded rule input vector, indicating which task is to be performed, which we denote $\mathbf{u}^\text{task-rule}$(``task name''); and second, a fixation input, a decrease of which is treated a `go' signal for the RNN to provide the appropriate output, which we denote $u^\text{fix}_t$. 

The expected output for each task is a response direction, which is again encoded in a ring of output units (which could represent, for example, a reach or saccade direction). These networks are trained using supervised learning with a cross entropy loss where the supervised target is a one-hot vector $\mathbf{y}$ representing the required location of the bump on the ring, and the model's output probabilities are constructed through a soft-max of the outputs, $\mathbf{o}_t$.
\begin{equation}
    L_{CE}(\mathbf{o}_t, \mathbf{y}) = -\sum_{i}\mathbf{y}_i\log \frac{e^{\mathbf{o}_{t,i}}}{\sum_{j}e^{\mathbf{o}_{t,j}}} 
\end{equation}
A schematic of the setup of the RNN is shown in Fig. \ref{fig:Fig1}c.
For each trial of each task, the inputs are constructed as a gaussian bump on each of the two rings whose mean is drawn independently 
 from a uniform distribution on the rings.

We compare LM-RNNs with different values of $d$ against fully-connected RNNs with the same number of neurons (and hence many more parameters; cf. Fig. \ref{fig:Fig2}a) and fully-connected RNNs of a smaller size but with the same number of parameters (cf. Fig. \ref{fig:Fig2}b). In all cases, LM-RNNs for \emph{any} value of $d$ are far more sample-efficient and learn the tasks with same asymptotic performance as compared to the fully-connected counterparts.
We also compare the performance of these models at the same fixed number of gradient steps early in training to show sample efficiency differences (Fig. \ref{fig:Fig2}a, \textit{inset}). 
We hypothesize that this increased efficiency for LM-RNNs may be due to the ability to use a high-dimensional computational space, while having a significantly smaller number of parameters to be learned, resulting in the faster training observed at a similar level of performance.

\subsection{`\yangtasks' are rapidly learned with simple autapse networks}

\begin{figure}
    \centering
    \includegraphics[scale = 0.22]{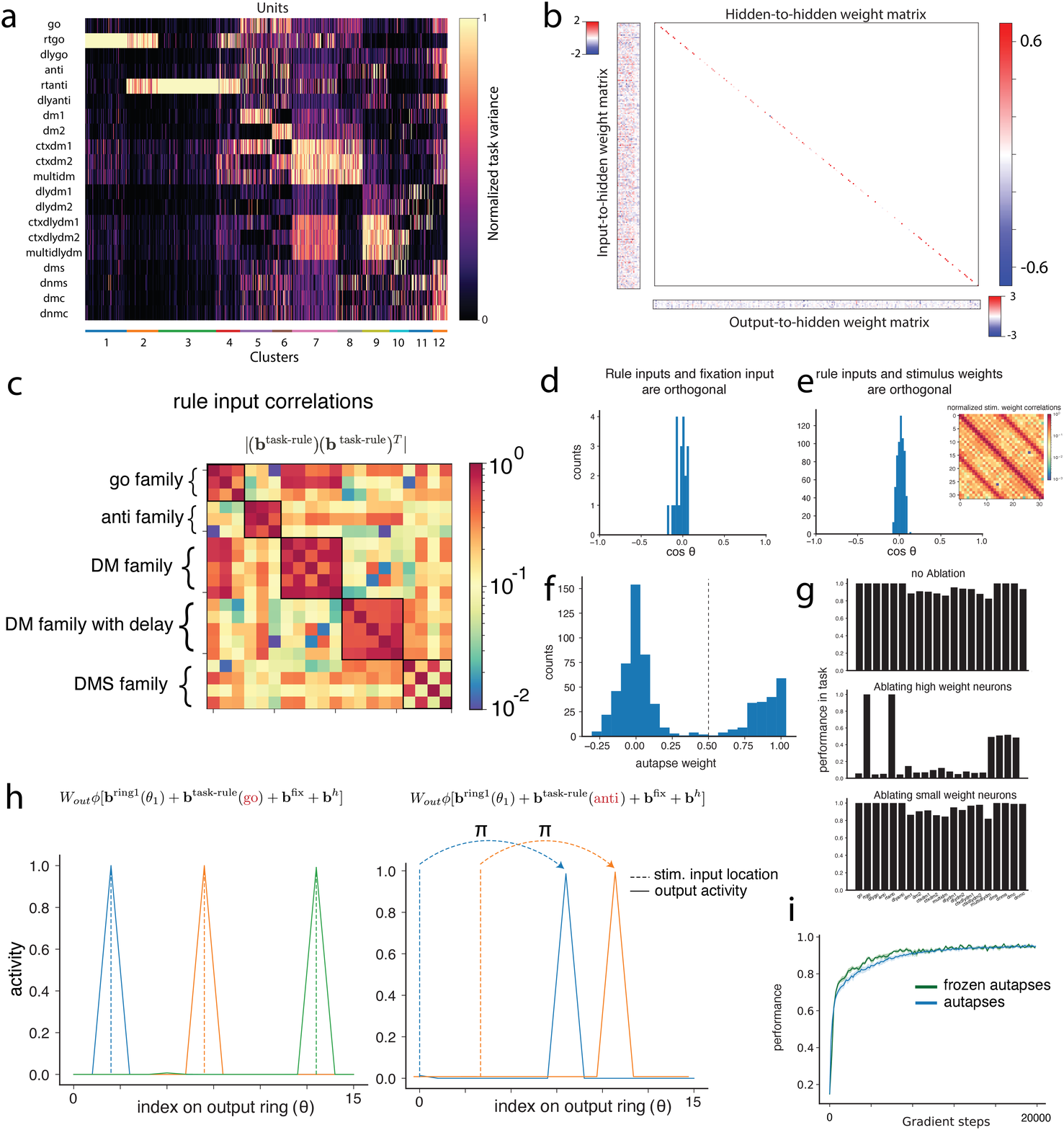}
    \caption{{\bf A suite of 20 cognitive tasks induces apparent modularity but is equally well-learned by a network of autapses (extreme spatial masking).} (a) An autapse network from Fig.\ref{fig:Fig1} trained on the original \yangtasks showing the formation of 12 specialized clusters; (b) The weights of model including the diagonal hidden to hidden matrix which clearly show the existence of no modular structure. (c) Normalized rule-input weight correlations show block structure that is consistent with the \yangtasks family structure. (d) Rule input weights are orthogonal to fixation input weights (e) Rule input weights are orthogonal to stimulus input weights and (inset) stimulus input weight correlations show a circulant structure. (f) A histogram of autapse weights shows the existence of 2 clusters of autapses, those with autapse weight close to 1 and those with autapse weight close to 0. (g) Ablating the small weight autapses results in no loss of task performance while ablating the high weight autapses reduces all task performances to random chance, apart from the 2 \texttt{reaction-time} tasks \texttt{rt-go} and \texttt{rt-anti} which do not need memory. (h) Explicitly adding the appropriate input vectors is sufficient to solve the tasks without the need for dynamics: (left) for task \texttt{go} and (right) for task \texttt{anti}. (i) A pool of frozen autapses can learn the \yangtasks faster than a pool of autapses and reach the same asymptotic performance.}
    \label{fig:SI_autapse}
\end{figure}

While we demonstrated that LM-RNNs at all $d$ perform better than fully-connected networks, we particularly note that $d=0$, (i.e., a `network' where each node is only connected to itself; in this case the network is simply a pool of disconnected autapses, and $\mathbf{G}$ is an identity matrix) also performs better than a fully-connected network in terms of learning speed while reaching the \emph{same asymptotic performance as a larger fully connected network.} Remarkably, as seen in Fig. \ref{fig:SI_autapse}, this pool of autapses continues to show `modularity' in the network through the nodal-task-variance based metrics similar to the results of fully connected RNNs in  \cite{yang2019task} --- however this apparent modularity clearly cannot be a result of any modular structures in the network due to the absence of any inter-node network connections. The autapse networks have no lateral connectivity and thus no way to share and reuse subtask structure across neurons. This suggests that such a task variance metric may simply be reflecting correlations between common inputs and similar input weights to hidden neurons. We verify this hypothesis in Fig. \ref{fig:SI_autapse}c where we plot the correlation between projection of the rule inputs to the hidden nodes, i.e., $\mathbf{b}^\text{task-rule}\text{(``task name'')} = \mathbf{W}^\text{task-rule}\mathbf{u}^\text{task-rule}\text{(``task name'')}$ where $\mathbf{W}^\text{task-rule}$ is the submatrix of input weights formed by the rule-input appropriate columns of $\mathbf{W}_{in}$. Since the $\mathbf{u}^\text{task-rule}$ is presented as simply one-hot encoded vectors, the correlation of the rule inputs is simply $[\mathbf{W}^\text{task-rule}]^T\mathbf{W}^\text{task-rule}$. We observe that this correlation matrix of input projections in itself appears to cluster corresponding to the common subtask structure of \yangtasks.

We thus hope that our results motivate the study of better metrics and techniques to inspect functional modularity in RNNs, which we leave for future work.

While these results are in themselves indicative of the advantages conferred by LM-RNNs, we note that the \yangtasks are evidently too simplistic to make any strong claims, since they do not even require a network of connected neurons to accomplish the task. We present here first a simplified analysis demonstrating how a pool of disconnected autapses can solve \yangtasks, and thereafter present a more rigorous battery of cognitive tasks to more robustly demonstrate the utility of LM-RNNs.

\subsection{Mechanistic analysis of the pool of autapses: A game of vector addition}
For mechanistic interpretability, we first examine the dynamics of a single \emph{linear} autapse in the presence of an input $b_t = \mathbf{W}_{in}\mathbf{u}_t$,
\begin{equation}\label{eq:autapse_sum}
    h_{t+1} = W h_t + b_t.
\end{equation}
This gives
\begin{equation}
    h_t = W^t h_0 + \sum_{n=0}^t W^{t-n} b_n.
\end{equation}
For a constant input, and for $0\leq W<1$ this can be simplified to give 
\begin{equation*}
    h_t = \left[h_0 - \frac{b}{1-W} \right]W^t + \frac{b}{1-W} 
\end{equation*}

Hence, we see that the autapse weight $W$ defines an effective timescale for the autapse dynamics. Over the weight-dependent timescale $\tau = -\frac{1}{\log W}$ the autapse relaxes to a fixed point given by $h^* = \frac{b}{1-W}$. Thus, autapses with low weight ($W \approx 0$) very quickly converge to their associated bias dependent fixed point given by $h^* \approx b$  while autapses with high weight ($W = 1$) effectively function as perfect memory, retaining their initial state $h_0$ and perform addition of the network inputs as they are received: \begin{equation}\label{eq:autapse_sum_weight_one}
     h_t = h_0 + \sum_{n=0}^t b_n. 
\end{equation}


Examining the weights of autapses in the trained autapse pool reveals that the weights appear to be seprable into two classes that correspond to the described above (Fig. \ref{fig:SI_autapse}f) --- those with weights close to zero, that would be expected to rapidly approach a fixed point; and those with weights close to one, that would be expected to perform vector addition.  We also observe that the rule inputs are approximately orthogonal to the fixation and ring inputs, Fig.\ref{fig:SI_autapse}d,e, suggesting that the dynamics due to the rule inputs act in a subspace that is independent from the fixation and input stimuli.



We hypothesized that performing vector addition would be sufficient to be able to solve the \yangtasks. We verified this in two ways: firstly, we found that ablating autapses with lower weights resulted in no loss in performance, whereas ablating the larger weight synapses lead to significant performance deficits reducing the performance on almost all tasks to chance \footnote{The chance performance for the delay match to sample family of tasks \texttt{dms,dnms,dmc,dnmc} is 50\% due to the way samples are drawn: half of the trials are matching and the other half are non-matching, refer to methods of \cite{yang2019task}. The performance of reaction-time tasks \texttt{rt-go} and \texttt{rt-anti} is not affected since they do not require memory.}, Fig. \ref{fig:SI_autapse}g; secondly, we found that explicit addition of the $\mathbf{b}^\text{task-rule}$(``task rule'') vectors to the inputs from the other modalities was sufficient to generate outputs that performed the task, Fig. \ref{fig:SI_autapse}h.



\subsubsection{A frozen pool of autapses}
Motivated by these ablation studies, we trained an autapse pool with untrainable fixed autapse weights set to 1. We refer to this pool as the ``frozen autapse pool". This setup also learns all 20 tasks notably faster than a regular autapse pool, Fig. \ref{fig:SI_autapse}i. In this case the autapses are effectively ``copy gates", storing perfect memory of their previous input and adding them to the input vector currently being received (cf. equation \ref{eq:autapse_sum_weight_one}). 


\subsection{\taskname: An expanded battery of cognitive tasks}

\begin{figure}
    \centering
    \includegraphics[width = \textwidth]{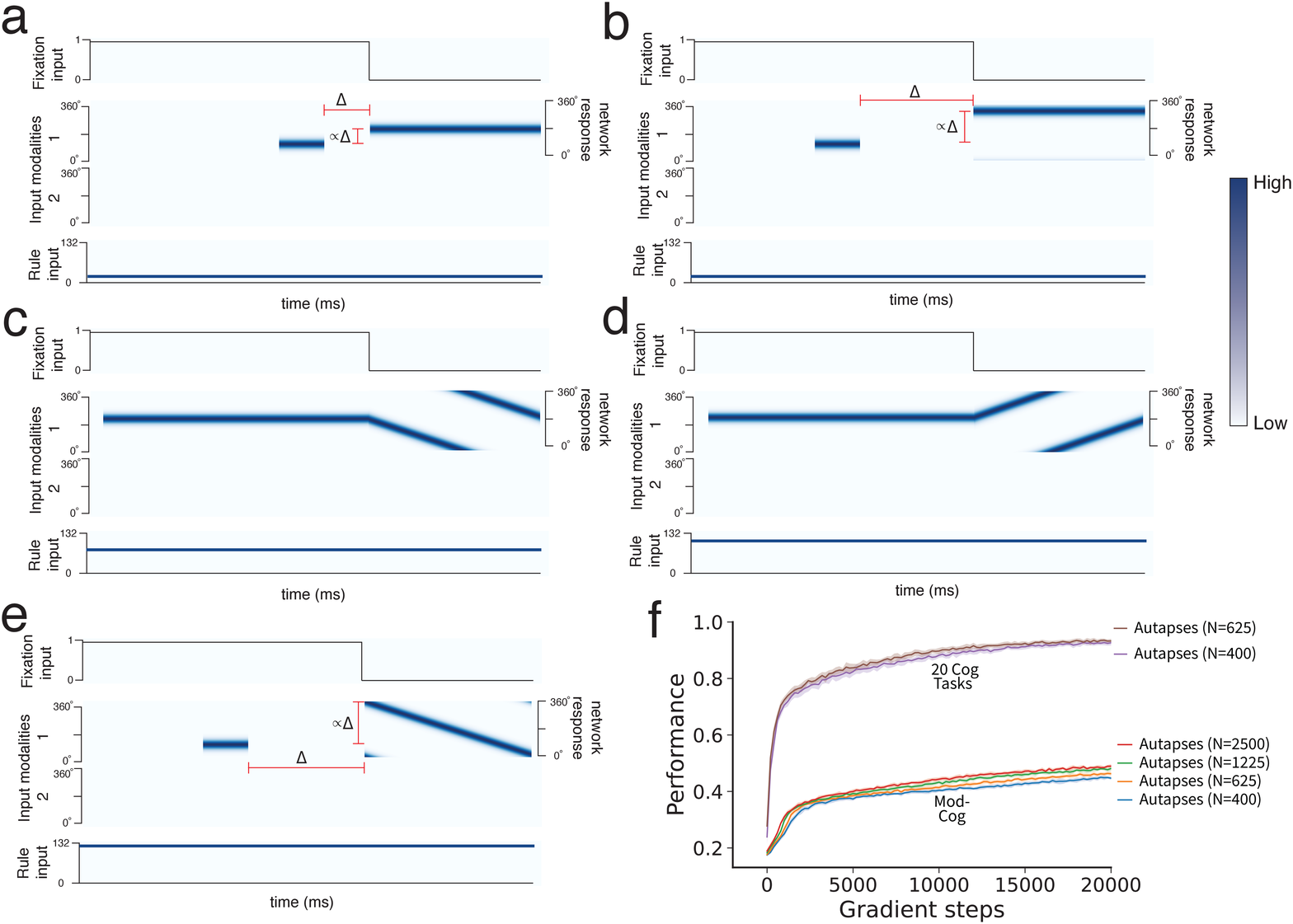}
    \caption{{\bf \taskname: A more-complex suite of upto 132 cognitive tasks.} Schematic of the tasks in \taskname, that extend the \yangtasks. (a,b) Tasks \texttt{DlyGo\_IntL} and \texttt{DlyGo\_IntL} as interval estimation extensions to the \texttt{DlyGo} task from \yangtasks for two different lengths of delay periods. (c,d) Tasks \texttt{Go\_SeqR} and \texttt{Go\_SeqL} as sequence generation extensions to the \texttt{Go} task from \yangtasks. (e) The compositional extension \texttt{DlyGo\_IntL\_SeqR} that combines both interval estimation as well as sequence generation. (f) \taskname introduces significantly more complex tasks as compared to \yangtasks, which can no longer be solved by a pool of autapses, regardless of their number.}
    \label{fig:Fig3}
\end{figure}

The reason the \yangtasks were trivially solved by a pool of (frozen) autapses is that the tasks were static, involving memory and fixed points but no dynamical computation like integration and sequence generation. These computations involve the manipulation of the information held in working memory. A circuit with integration properties is able in principle to powerfully generalize across tasks in a way that networks that exhibit the same set of states only as stable fixed points are not able to \cite{klukas2020efficient, khona2022attractor} . To perform a more robust demonstration of the utility of LM-RNNs, we construct a battery of new tasks based on the principle of integration, and related to cognitive science problems such as interval estimation, physical and mental navigation, and sequence generation. 
We build a set of modular and compositionally constructed tasks, using the \texttt{neurogym} framework \cite{molano2022neurogym}, which in turn was built on the AI opengym environment. 

In particular, we build extensions, that incorporate additional complexity in two main forms: integration and sequence generation.

In the case of integration, we consider the set of `delay' based tasks in the \yangtasks. In the \yangtasks, 12 out of 20 tasks involve a delay period in the presented input, wherein the network is expected to persistently hold the input presented in an internal memory before performing a task-relevant computation. To incorporate interval estimation in these tasks, we require the output to be discplaced with respect to the orignally expected output by a magnitude dependent on the length of the delay period. To this end, we choose the delay length randomly from a uniform distribution (as opposed to the fixed delay length considered in \yangtasks). As representative examples, in Fig. \ref{fig:Fig3}a,b we show the inputs and expected outputs for two different delay period lengths in the \texttt{DlyGo\_IntL} task, constructed as an interval estimation extension to the \texttt{DlyGo} task from \yangtasks.
For each of the 12 tasks, the interval-dependent-discplacement may be either of clockwise or anti-clockwise, resulting in the introduction of 24 new tasks.

In the case of sequence generation, the output of each task is modified to not be a single static direction, but instead a time-varying output corresponding to a drifting direction starting at a particular point (dependent on the particular task). The direction of drift can be changed dependent on the particular task. As representative examples, in Fig. \ref{fig:Fig3}c,d we show the inputs and expected outputs for the \texttt{Go\_SeqL} and \texttt{Go\_SeqR} tasks, constructed as an sequence generation extensions to the \texttt{Go} task from \yangtasks.
This introduces 40 new tasks based on the earlier set of 20 tasks, with the output of each task drifting either clockwise or anti-clockwise.

This completes the construction of the 64 new tasks that we use in conjunction with the original 20 tasks as the tasks used for our main set of results hereafter in this paper. We refer to this set of 84 tasks as \taskname. 
We note however that our modifications to the tasks are modular in nature (which is similar in spirit to the already existing modular subtask structure in the \yangtasks). This allows for an additional extension of 48 more tasks that may be generated by a composition of the sequence generation and interval estimation extensions (such as the \texttt{DlyGo\_IntL\_SeqR} task shown in Fig. \ref{fig:Fig3}e). For simplicity, we do not use these additional 48 tasks in our main results; however they are included in the repository of tasks that we provide at github link (will be inserted upon acceptance; provided as \texttt{.zip} file in supplementary material).

The rule input used for \taskname is encoded as a one-hot vector, similar to the setup used in \cite{yang2019task}. This ensures that, by construction, the rule input cannot be directly used as a signal to help decompose tasks into having common subtasks.

To demonstrate that \taskname is significantly harder than \yangtasks, we demonstrate in Fig. \ref{fig:Fig3}f that a pool of autapses is incapable of acheiving significant performance levels, in sharp contrast with the \yangtasks. Moreover, this is independent of the number of autapses --- even pools that are $\sim 6$ times larger than those necessary for solving \yangtasks are unable to produce larger than 50\% performance accuracy.

\subsection{RNN and LM-RNN performance on \taskname}

\begin{figure}[t]
    \centering
    \includegraphics[width = \textwidth]{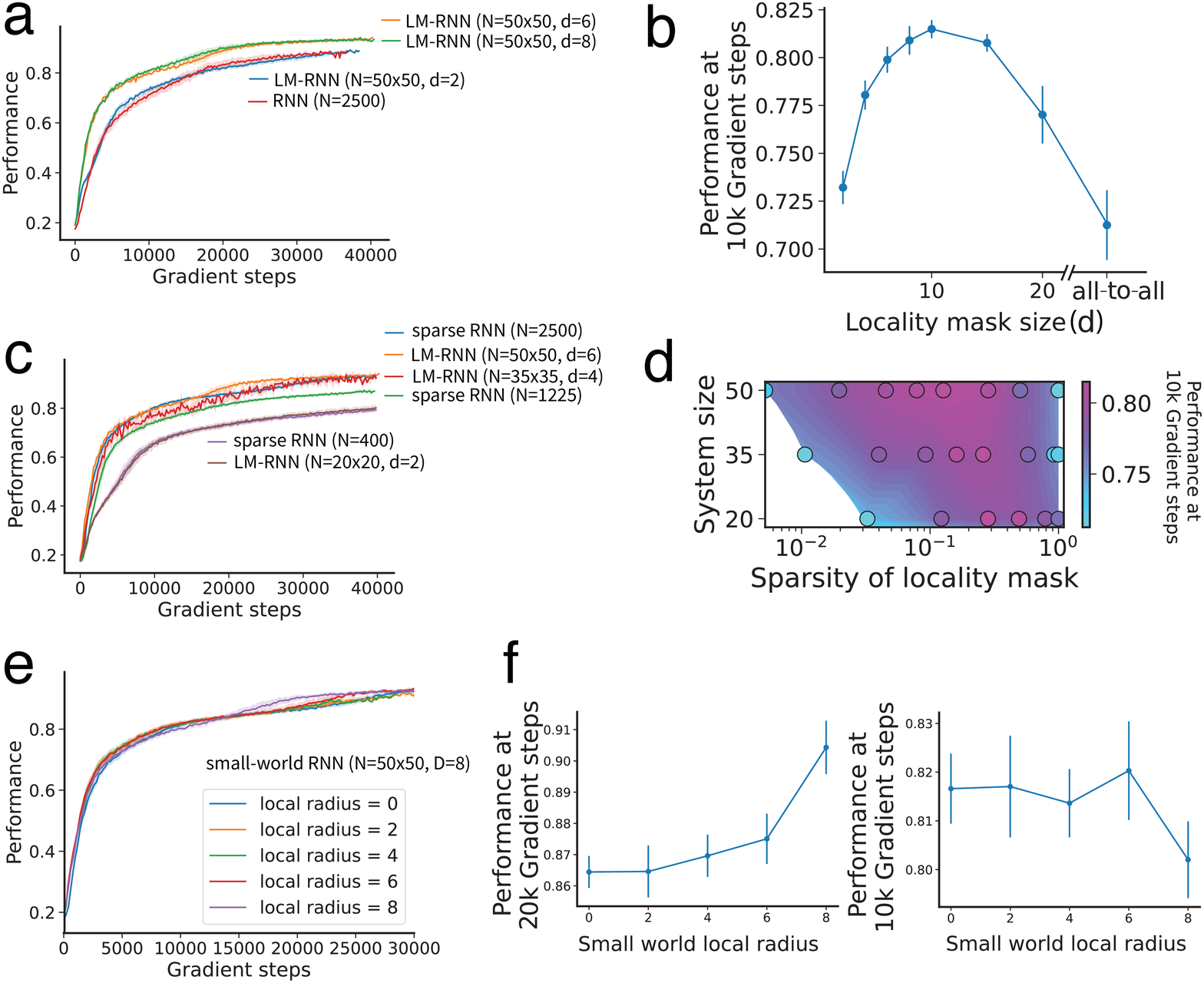}
    \caption{{\bf Faster learning and better asymptotic performance of LM-RNNs and sparse RNNs on \taskname} (a) LM-RNNs for all values of $d$ perform equally well or better than a fully-connected network with the same number of nodes. This improvement takes the form of faster learning as well as better asymptotic performance. (b) Network performance at $10^4$ gradient steps for different values of $d$ and networks of size $50\times 50$, demonstrating an optimal locality size that leads to the fastest learning. (c) LM-RNN performance for varying system size for a fixed network sparsity of $\sim$4\%, as compared with fixed random sparse networks with the same sparsity. Performance for sparse RNNs is similar to or only slightly worse than LM-RNNs with the same sparsity (d) Network performance as a function of system size ($N^2=\text{System Size}$) and sparsity of the locality mask (i.e., the fraction of nonzero entries in the hidden-to-hidden weight matrix). At larger system sizes, the optimal sparsity for best performance is lower. (d) (e) }
    \label{fig:Fig4}
\end{figure}

As earlier, we examine the performance of LM-RNNs with varying $d$ on \taskname, and compare them with fully-connected RNNs as a function of the number of gradient steps in training (cf. Fig. \ref{fig:Fig4}a). Here again we see that LM-RNNs, for appropriately chosen values of $d$ train significantly more rapidly as compared to fully connected networks. Due to the increased task complexity (as evidenced by Fig. \ref{fig:Fig3}f), locality masks corresponding to very small values of $d$ result in suboptimal performance (which is nonetheless similar to the $d\to\infty$ corresponding to the fully connected network). Instead, intermediate small values $d$ outperform all other values of $d$, as shown in Fig. \ref{fig:Fig4}b, indicating an optimal nontrivial locality mask $d$. 
For smaller networks, the optimal value of $d$ corresponds to increasingly larger fractions of all edges in the network (cf. Fig. \ref{fig:Fig4}d), indicating that the optimal may depend more directly on the complexity of the tasks to be solved, rather than scaling with the number of nodes in the network.

For most LM-RNNs, a \emph{random} sparse graph with the same number of incoming edges from each node performs almost as well as the graph chosen through locality masking, as shown in Fig. \ref{fig:Fig4}c. Nevertheless, in each case the locality mask performs as well as or slightly better than the random sparse graph, while performing significantly better than dense fully-connected networks. For the case of small values of $d$, we hypothesize that the slight improvement of locality masks may arise from the presence of disconnected components in random sparse graphs in networks with low degrees. For clustering visualizations based on nodal task variance \cite{yang2019task}, see Appendix Figs. \ref{fig:clustern20d2},\ref{fig:clustersn20d100},\ref{fig:clustersn20d6},\ref{fig:clustersn50d10} and \ref{fig:clustersn50d100}. 

We also examined small-world networks\cite{watts1998collective} as graphs that interpolate between random sparse graphs and the locality masks of LM-RNNs. In particular, we considered graphs with a fixed number of total synapses corresponding to an LM-RNN with $d=8$. Small-world graphs were constructed with this number of synapses: connections were formed between nodes within a `local radius' distance of each other, and random sparse connections were added until the target total number of synapses. A local radius of zero then corresponds to a completely random sparse graph, and a local radius of 8 corresponds to the LM-RNN that we have been considering thus far. 
We observed that there was no significant variation of performance across different local radi at $10^3$ gradient steps, however, at $2\times10^3$ gradient steps the LM-RNN appears to perform slightly  better than any small world network, Fig. \ref{fig:Fig4}f (left).

\subsection{Controls -- comparisons with various baselines for sparsity} 
We perform comparisons across 3 baselines. First, fully-connected networks with the same number of nodes; second, fully-connected networks with the same number of synapses; and third, fully-connected networks with the same number of nodes, but with an additional $L_1$ regularization term in the loss function to promote the discovery of sparse solutions through training. As we demonstrate in Table \ref{tab:Tab1}, LM-RNNs reach a higher performance earlier than all other comparable models. While sparse networks acheived through $L_1$ regularized models perform better than fully-connected models at some choices of regularization strengths, they acheive this better performance slower than LM-RNNs with static sparsity. Thus while sparsity is clearly benefecial to improved performance in the multitask setting of \taskname, choosing this sparsity in a fixed, task-independent fashion at the start of training results in faster training.

\begin{table}[]
    \centering
    \begin{tabular}{|l||c|c|c|}
    \hline
         & Performance at  & No. of nonzero & No. of nonzero \\
         & 10k gradient steps & weights & hidden-to-hidden weights \\
         \hline 
        \textbf{LM-RNN($\mathbf{N=50 \times 50, d=6}$)} &  $\mathbf{0.799 \pm 0.007}$ & $\mathbf{6.17\times 10^5}$ & $\mathbf{2.82\times 10^5}$\\ 
        \textbf{LM-RNN($\mathbf{N=50 \times 50, d=10}$)}& $\mathbf{0.815 \pm 0.005}$ & $\mathbf{1.13\times10^6}$  & $\mathbf{7.92\times10^5}$\\ 
        RNN($N=676$)           & $0.69 \pm 0.01$   & $5.47\times 10^5$ & $4.57\times10^5$\\
        RNN($N=2500$)           & $0.71 \pm 0.02$   & $6.58\times 10^6$ & $6.25\times 10^6$\\
        $L_1$ RNN($N=2500$), $\lambda=10^{-4}$ & $0.59 \pm 0.02$  & $4.00\times10^5$  & $6.51\times10^4$ \\
        $L_1$ RNN($N=2500$), $\lambda=10^{-5}$ & $0.72\pm 0.01$   & $5.16\times 10^5$ & $1.81\times10^5$ \\
        $L_1$ RNN($N=2500$), $\lambda=10^{-6}$ & $0.73\pm 0.02$   & $1.02\times10^6$  & $6.87\times10^5$\\
        $L_1$ RNN($N=2500$), $\lambda=10^{-7}$ & $0.71\pm 0.01$   & $1.86\times10^6$  & $1.52 \times 10^6$\\
        \hline
    \end{tabular}
    \caption{{\bf Comparison across networks with similar number of trainable parameters and nonzero weights.} $L_1$ RNNs are fully-connected RNNs with an $L_1$ regularization to promote sparse solutions. In this case, the number of nonzero weights is counted as the number of weights above a small threshold in the trained RNN. See Appendix Fig.\ref{fig:SI_perf_curves_for_sparse} for learning curves}
    \label{tab:Tab1}
\end{table}


%% file: 03_discussion.tex
\section{Discussion}

Through our results, we have demonstrated that a simple fixed sparse graph can provide a large increase in the sample-efficiency of the training process in single-layer recurrent networks. These results could in principle be extended to other architectures and tasks but we restricted our experiments to simple RNNs and cognitive tasks for their relevance to systems and computational neuroscience.

As RNNs trained on multitask learning problems become more popular as models of PFC and other associated brain regions \cite{driscoll2022flexible,hummos2022thalamus,riveland2022neural,elucidating20,duncker20organizing,masse2022rapid,kao2021natural}, we need to be more cognizant about how the computational complexity of the task set used affects learnt representations. By demonstrating that a pool of unconnected autapses can solve the \yangtasks \cite{yang2019task}, we have shown that this apparently complex set of tasks can be solved without any communication between neurons. This raises the hypothesis that appropriately tuned input weights corresponding to the ``rule inputs" along with static memory are enough to solve many tasks. 
Here we constructed \taskname as a set of tasks with increased complexity by adding more modular subcomponents that require the RNN to perform non-trivial computations. Thus we provide a more reasonable multitask setting that leads to richer solutions and may be a better testbed for exploring shared motifs and representations across modular subtasks. 
Another possible direction for future study to increase task complexity for \yangtasks has been to eliminate rule inputs and force the network to infer the task needed to be solved. This would likely involve some flavor of predictive coding. A recent work \citet{hummos2022thalamus} makes progress in this direction. 

We have found that given a certain amount of task complexity and network size, there is an optimal amount of locality masking that provides the most benefit to learning, both in terms of sample efficiency and asymptotic performance. While we have shown this result for recurrent networks, qualitatively similar results on an optimal value of sparsity have been derived analytically for cerebellum-like feedforward architectures \cite{litwin2017optimal}. It remains an open question to theoretically investigate the relationship between task complexity and the amount of sparsity that is needed: if the network is too sparse, it will not have enough expressivity to perform well on the dataset, while if the network is too dense, the benefits provided by sparsity will not be exploited.


Although our matrices are extremely sparse and would be well suited to sparse matrix representations, we still maintain dense matrix dataypes for all of the training and evaluation processes since standard libraries like PyTorch do not have native support for these data structures. As sparse matrices get more common and their usefulness more apparent, as has been pointed out before  \cite{liu2021selfish}, it will be very useful for native deep learning software and hardware implementations on GPUs to exploit the potential efficiencies of very sparse matrix structures.

%% file: 07_acknowledgements.tex
\section{Acknowledgements}
We thank Guangyu Robert Yang for open-sourcing the 20 cognitive tasks set and the NeuroGym framework \cite{molano2022neurogym} released under an \texttt{MIT License}. This work was supported by the MathWorks Science Fellowship to MK, the Simons Foundation through the Simons Collaboration on the Global Brain, the ONR, and the Howard Hughes Medical Institute through the Faculty Scholars Program to IRF. 

%% file: 04_references.tex
\bibliographystyle{plainnat}
\bibliography{main.bib}

%% file: 06_appendix.tex
\appendix

\section{Methods and Hyperparameters}
PyTorch was used for all simulations. All networks were trained with supervised learning using a cross entropy loss. The optimizer used was Adam  \cite{kingma2014adam} with a learning rate of $10^{-3}$. Each trial was drawn randomly and independently and the tasks were randomly interleaved. The Neurogym  \cite{molano2022neurogym} environment was used to create tasks and the training data for the model.\\
For every performance curve, we trained 15 RNNs with different random seeds and used the averaged curve for plotting and computing the optimal locality mask sizes.\\
An expanded description of the \taskname tasks and how to create them will be made available at the Github repository after acceptance.

To measure the number of clusters that the hidden nodes of the RNN partition into to solve a task, we use first use agglomerative hierarchical clustering \footnote{https://docs.scipy.org/doc/scipy/reference/generated/scipy.cluster.hierarchy.linkage.html} to obtain a linkage tree for the variance of each node across different inputs for a given task. Then, the silhouette score  \cite{rousseeuw1987silhouettes} is evaluated for each possible linkage-based cluster partition and the partition with the highest score is selected to represent the clustering of the RNN.

Graph theoretic measures shown in Fig. \ref{fig:Fig1}d were computed using the Python package `\texttt{networkx}' version 2.8.4. 


\begin{figure}
    \centering
    \includegraphics[scale = 0.4]{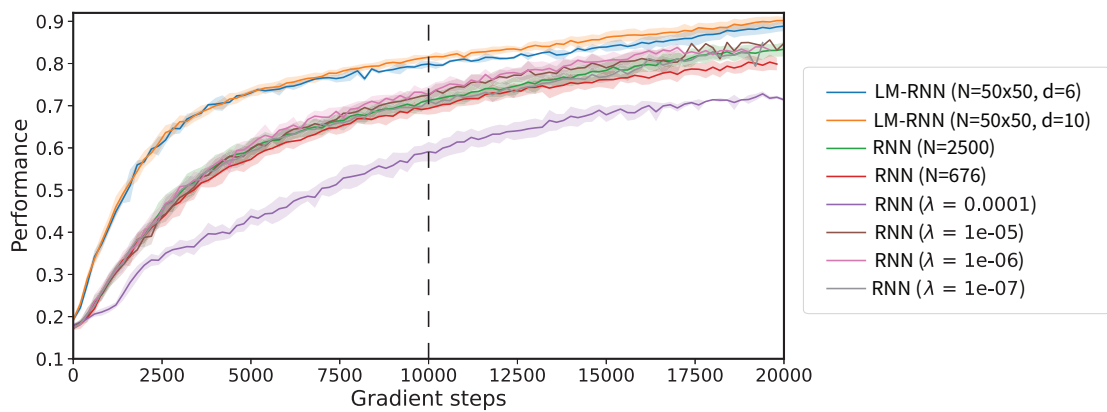}
    \caption{\textbf{Performance curves} of models trained with varying values of sparsity (L1) regularization showing that a fixed local mask (of similar sparsity, cf. Table \ref{tab:Tab1}) is better.}
    \label{fig:SI_perf_curves_for_sparse}
\end{figure}


\begin{figure}
    \centering
    \includegraphics[scale=0.15]{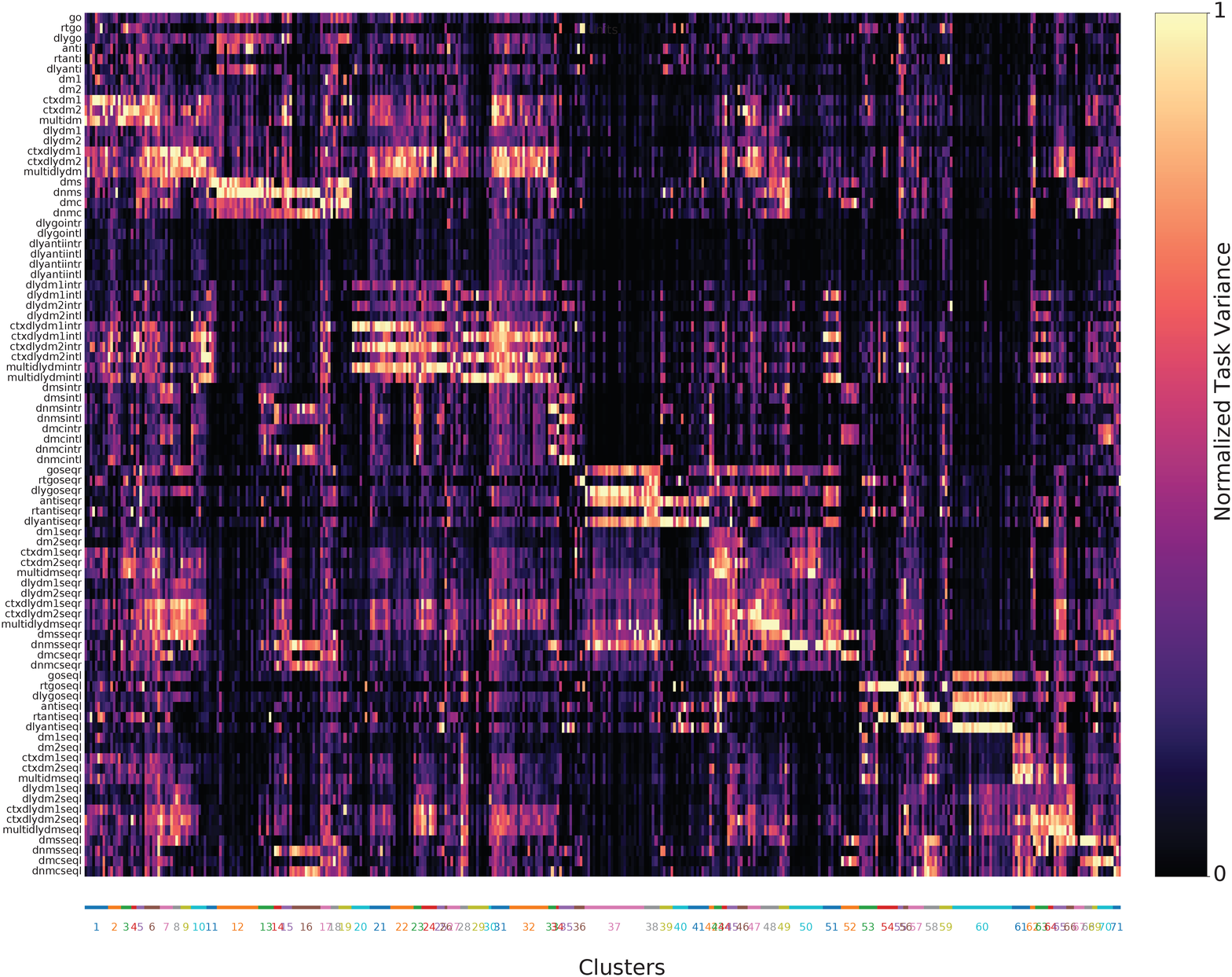}
    \caption{An LM-RNN(N=20x20,d=2) trained on 84 \taskname tasks showing the formation of clusters }
    \label{fig:clustern20d2}
\end{figure}

\begin{figure}
    \centering
    \includegraphics[scale = 0.15]{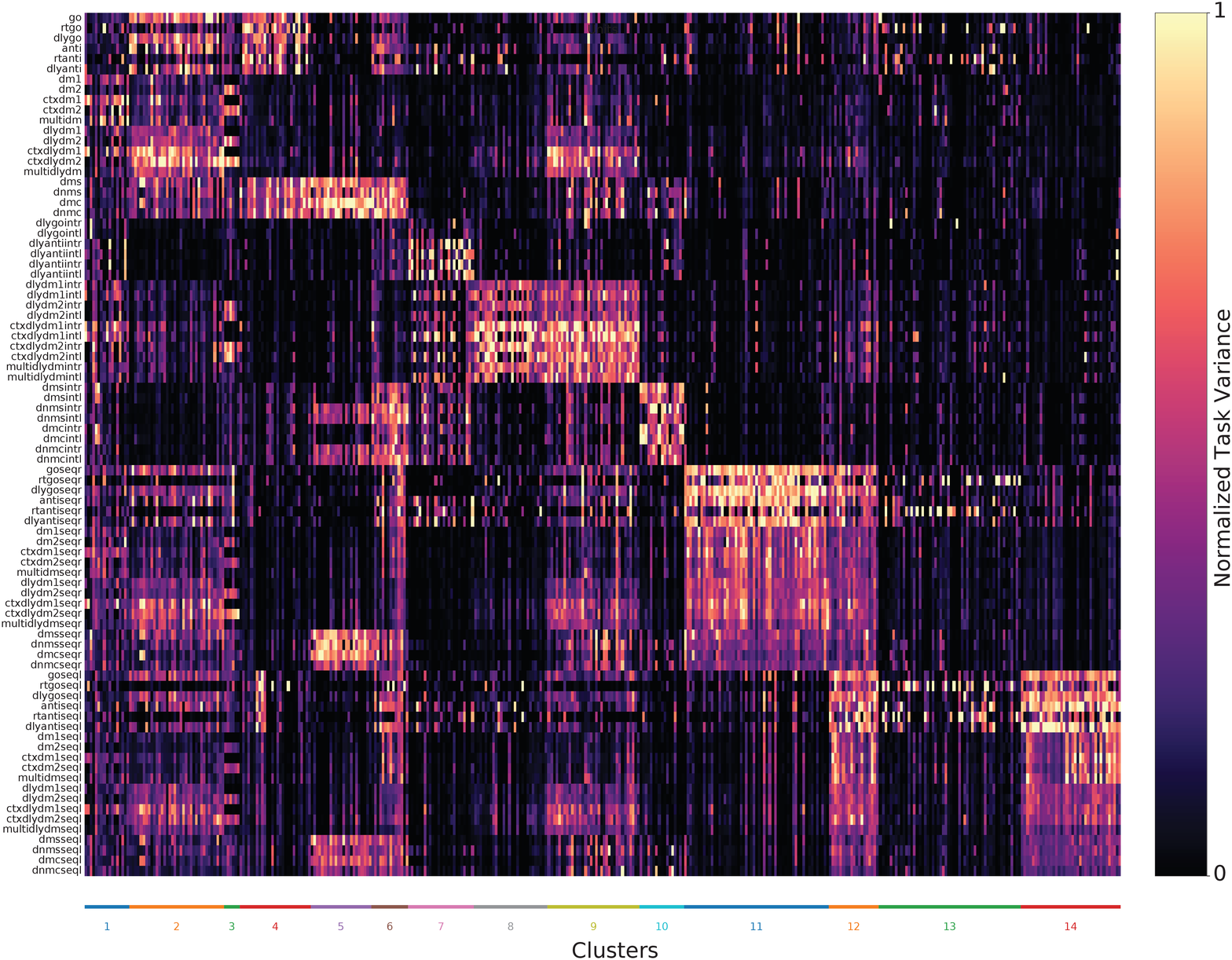}
    \caption{An LM-RNN(N=20x20,d=6) trained on 84 \taskname tasks showing the formation of clusters }
    \label{fig:clustersn20d6}
\end{figure}

\begin{figure}
    \centering
    \includegraphics[scale = 0.15]{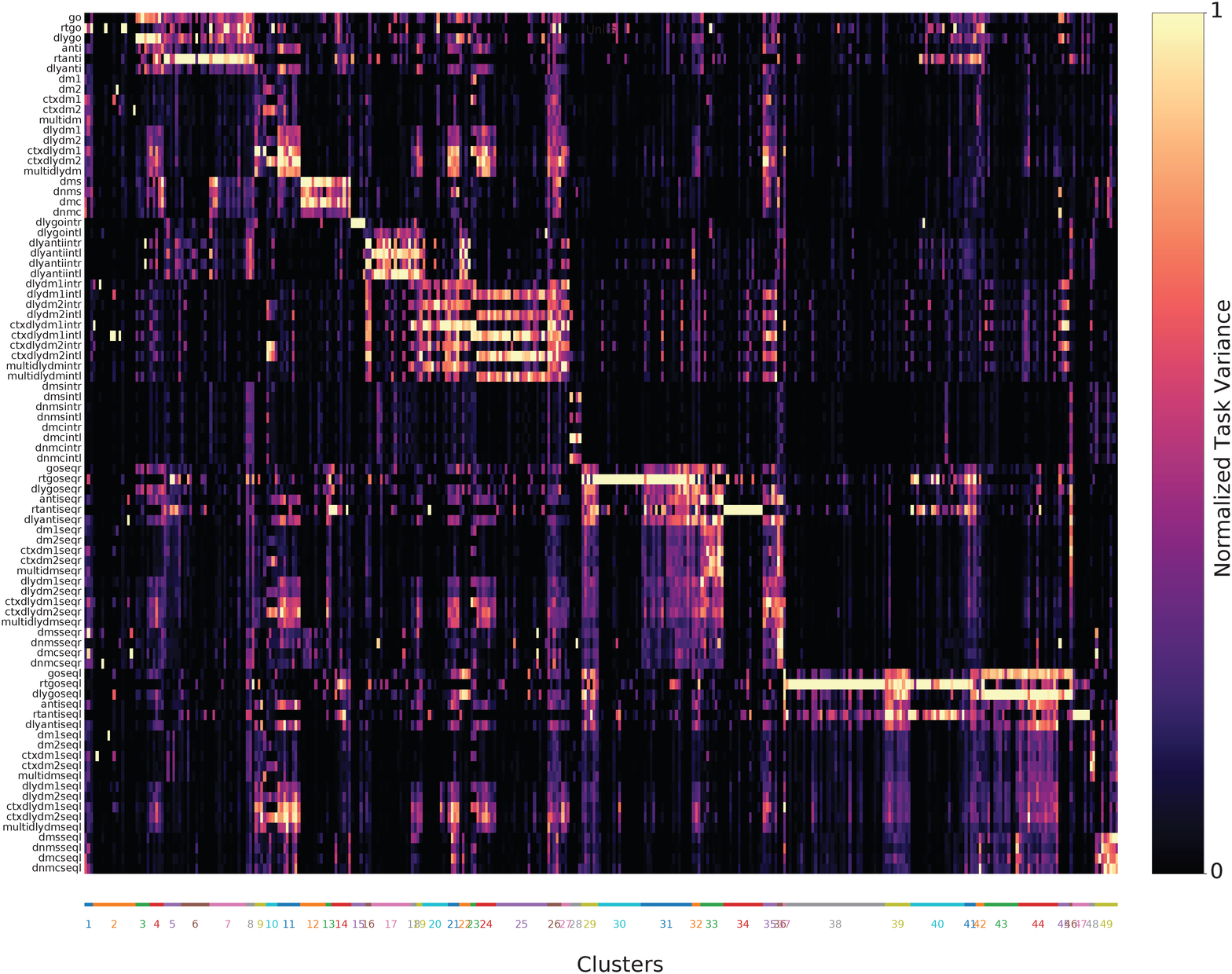}
    \caption{An RNN(N=400) trained on 84 \taskname tasks showing the formation of clusters }
    \label{fig:clustersn20d100}
\end{figure}

\begin{figure}
    \centering
    \includegraphics[scale = 0.22]{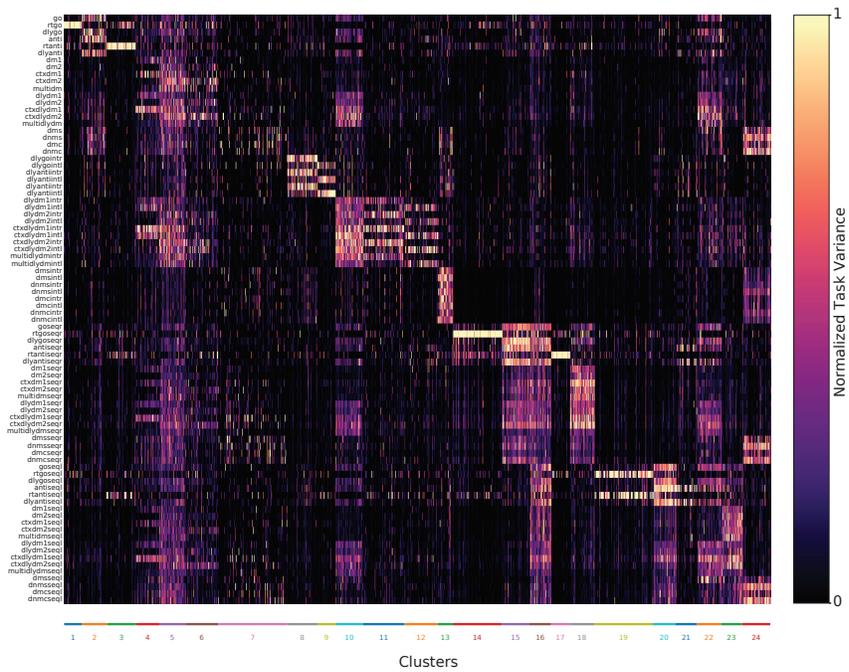}
    \caption{24 Clusters formed in an LM-RNN(N = 50 $\times$ 50, d = 10) trained on 84 \taskname tasks.}
    \label{fig:clustersn50d10}
\end{figure}

\begin{figure}
    \centering
    \includegraphics[scale = 0.22]{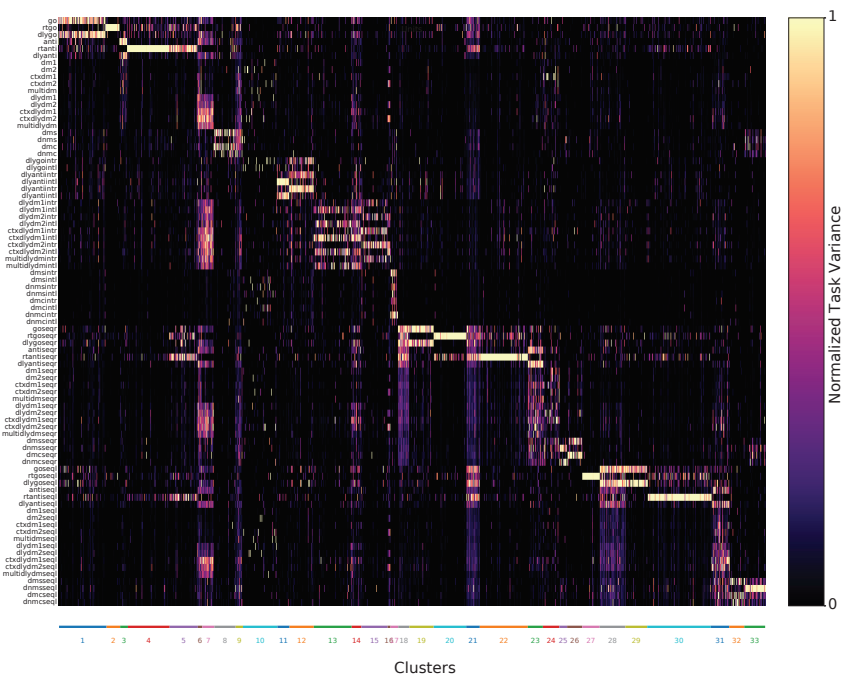}
    \caption{33 Clusters formed in an RNN(N=2500) trained on 84 \taskname tasks, contrast with previous figure where the number of clusters is smaller.}
    \label{fig:clustersn50d100}
\end{figure}